\documentclass{elsart1p}
\usepackage{graphicx}
\usepackage{amssymb}
\begin{document}
\begin{frontmatter}
%
%
%
\title{The ABC Effect in Double-Pionic Nuclear Fusion and a $pn$ Resonance as its Possible Origin}
%
%
\author{M. Bashkanov for the CELSIUS/WASA and WASA-at-COSY Collaborations}
\address{Physikalisches Institut, Universit\"at T\"ubingen, Germany}
\begin{abstract}
The ABC effect -- a long-standing puzzle in double-pionic fusion -- has been
reexamined by the first exclusive and kinematically complete measurements of
solid statistics for the fusion reactions $pn \to d\pi^0\pi^0$, $pd \to
^3$He$\pi\pi$ and $dd \to ^4$He$\pi\pi$ using the WASA detector, first at
CELSIUS and recently at COSY --- the latter with a statistics increased
by another two orders of magnitude. In all cases we observe a huge low-mass
enhancement in the $\pi\pi$-invariant mass accompanied by a pronounced
$\Delta\Delta$ excitation. For the most basic fusion reaction, the $pn \to
d\pi^0\pi^0$ reaction, we observe in addition a very pronounced resonance-like
energy dependence in the total cross section with a maximum 90 MeV below the
$\Delta\Delta$ mass and a width of only 50 MeV, which is five times smaller
than expected from a conventional $t$-channel $\Delta\Delta$ excitation. This
reveals the ABC effect to be the consequence of a $s$-channel resonance with
the formfactor of this dibaryonic state being reflected in the low-mass
enhancement of the $\pi\pi$-invariant mass. From the fusion reactions to
$^3$He and $^4$He we learn that this resonance is robust enough to
survive even in nuclei.
\end{abstract}
\end{frontmatter}
%
\section{}
\label{}
The two-pion production in nucleon-nucleon collisions has been studied
systematically at the storage rings CELSIUS
and COSY. At energies near threshold the single-nucleon 
excitation is favored. Since single-$\Delta$ excitation is highly suppressed,
single-Roper excitation dominates the near-threshold $\pi\pi$ production. 
At higher energies double-nucleon excitations come into play. For $T_p >$ 1
GeV  the $\Delta\Delta$ 
excitation becomes the leading process as observed in the respective 
$N\pi$ invariant mass spectra. Whereas this is in qualitative agreement with
theoretical predictions, we find from the full information present in the data
that the $\Delta\Delta$ system behaves very differently 
from what has been predicted.

In order to shed more light onto this problem, a long-standing puzzle in
$\pi\pi$ production has been reexamined: the so-called ABC-effect. 
This acronym stands for an unexpected enhancement at low masses in the
invariant $\pi\pi$ mass spectrum $M_{\pi\pi}$ first observed by {\bf
  A}bashian, {\bf B}ooth and  {\bf C}rowe in the double 
pionic fusion of deuterons and protons to $^3$He \cite{abc}. Follow-up
experiments \cite{cod}
revealed this effect to be of isoscalar nature and to show up in cases where
the two-pion production process leads to a bound nuclear system. With
the exception of low-statistics bubble-chamber measurements all 
experiments conducted on this issue have been inclusive measurements carried
out preferentially with single-arm magnetic spectrographs for the detection
of the fused nuclei.

Initially the low-mass enhancement had been interpreted as an unusually large 
$\pi\pi$ scattering length and evidence for the $\sigma$ meson \cite{abc},
respectively. Since the effect showed up particularily clear at beam energies
corresponding to the excitation of two $\Delta$s in the nuclear system, the ABC
effect was interpreted later on by a $t$-channel $\Delta\Delta$
excitation in the course of the reaction process leading to both a low-mass
and a high-mass enhancement in isoscalar $M_{\pi\pi}$ spectra
\cite{ris,barn,anj,gar,mos,alv}.

\begin{figure} 
\centering
\includegraphics[width=0.68\columnwidth]{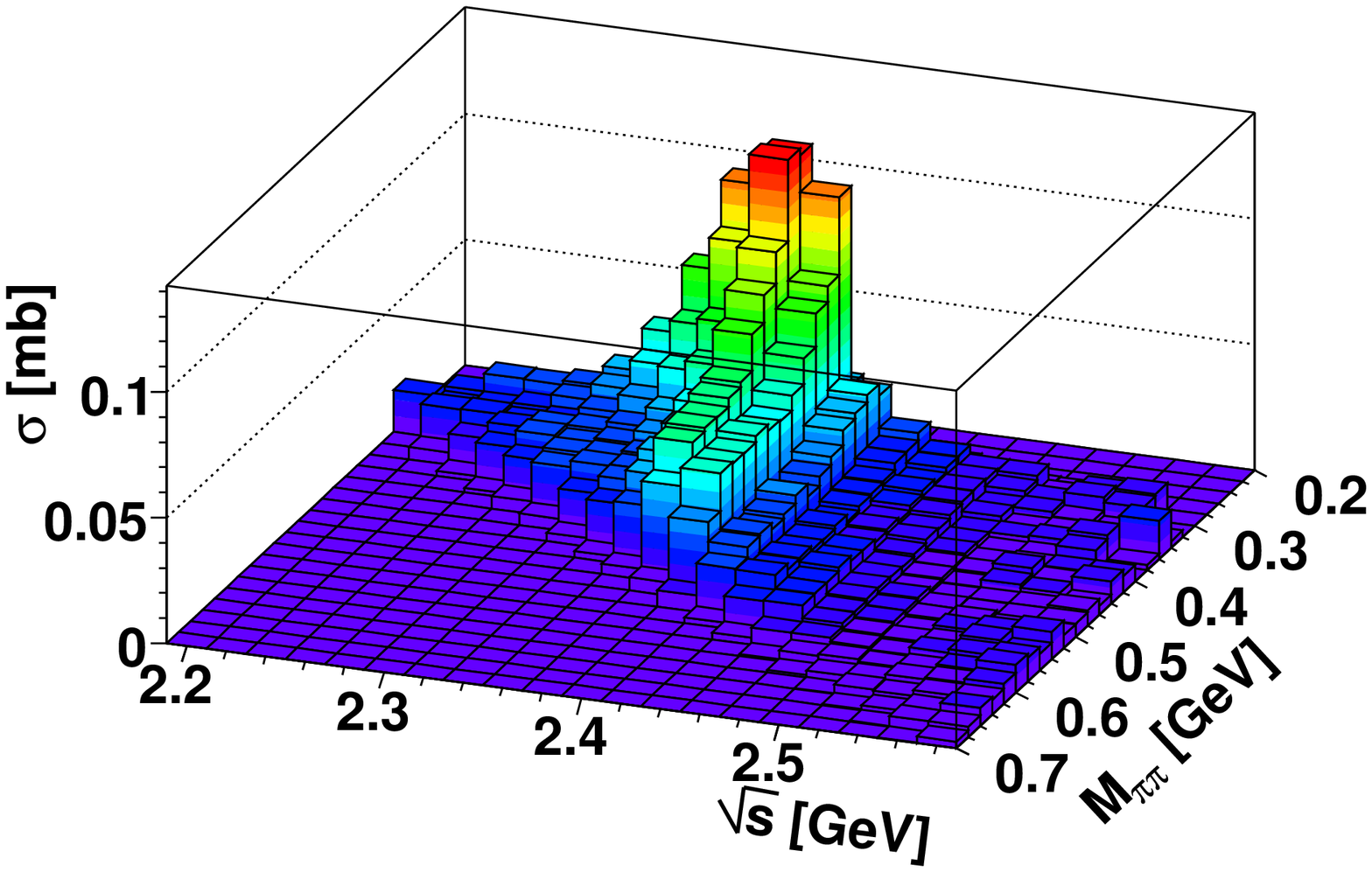}
\includegraphics[width=0.68\columnwidth]{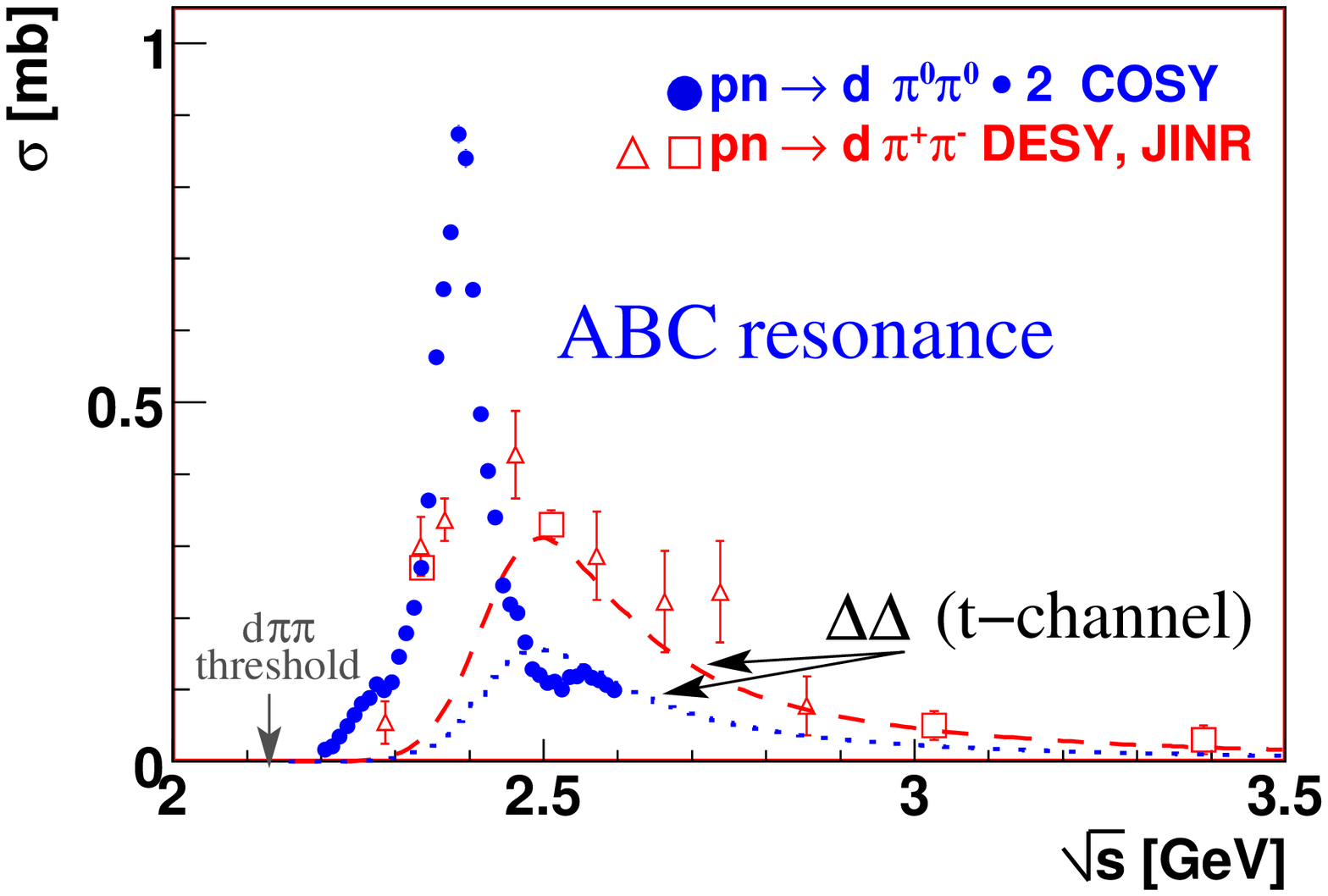}
\caption{Top: Energy dependence of the $\pi^0\pi^0$ invariant mass
  $M_{\pi^0\pi^0}$ depicted by a 3D-plot of $M_{\pi^0\pi^0}$ versus the total
  energy in the center-of-mass system $\sqrt{s}$.
  Bottom: Energy dependence of the total cross section for the 
  $pn\rightarrow 
  d\pi^+\pi^-$ reaction from threshold  up to
  $\sqrt{s}$ = 3.5 GeV. Data for the $d\pi^+\pi^-$ channel are from JINR Dubna
  (squares) and DESY (open triangles). The preliminary results
  of this work for 
  the $\pi^0\pi^0$ channel --- scaled by the isospin factor of 
   two --- are given by the full circles. Dashed and dotted lines
   represent $t$-channel $\Delta\Delta$ calculations for $\pi^+\pi^-$ and
   $\pi^0\pi^0$ channels, respectively.
}
\label{fig1}
\end{figure}


With the WASA $4\pi$ detector\cite{barg} first at CELSIUS and more recently in
the COSY ring we have measured the double-pionic fusion 
processes to D, $^3$He and $^4$He exclusively and kinematically complete
with a statistics which is orders of magnitude higher than  in
previous measurements. For the most basic fusion process, the one leading to
deuterium, first preliminary results from the data analysis
are now available, see Fig.~\ref{fig1}, which are in full support but of much
higher quality than those taken at CELSIUS \cite{MB}: On top the spectrum
of the $\pi^0\pi^0$-invariant mass $M_{\pi^0\pi^0}$ is shown in dependence of
total energy in the center-of-mass system. Our data exhibit an enormeous
low-mass enhancement, however, no apparent 
high-mass enhancement as predicted by conventional $\Delta\Delta$
calculations. Moreover the $\sqrt{s}$ dependence in the total cross section
(bottom) reveals an unexpected 
and narrow resonance-like structure --- again in contradiction to the
conventional $\Delta\Delta$ process (dotted curve for
the  $\pi^0\pi^0$ channel). As we see, the ABC-effect,
i.e. the low-mass enhancement is correlated to this narrow
resonance structure, which has its maximum 90 MeV below the  $\Delta\Delta$
mass and a width of only 50 MeV, {\it i.e.} five times smaller than expected
from the conventional 
$t$-channel $\Delta\Delta$ excitation. Indeed, describing this structure by a
$s$-channel {\it ansatz} leads to a surprisingly good
description of both the total and the differential distributions including the
ABC effect in the $M_{\pi^0\pi^0}$ spectra.  The quantum numbers of this
resonance have to be $I(J^P) = 0(1^+, 3^+)$. Such a 
resonance has been predicted by various theoretical calculations
\cite{ping,barnes,kam,oka,kuk,mot}, some of 
which even predict this resonance to be a member of a dibaryon multiplet
\cite{malt}. 

From the fact that the ABC effect is observed also for double-pionic fusion
processes to heavier nuclei, we conclude that this resonance is 
robust enough to survive in nuclei. 

This work has been supported by the BMBF(06TU261), COSY-FFE, 
DFG(Europ. Graduiertenkolleg 683) and the Swedish Research Council.
%
%
%

%
\end{document}